\documentclass[12pt]{article}
\begin{document}
\def\parno{\par \noindent}
\def\lr #1{\mathrel{#1\kern-.75em\raise1.75ex\hbox{$\leftrightarrow$}}}
\def\bi{\begin{itemize}}
\def\ei{\end{itemize}}
\def\benu{\begin{enumerate}}
\def\eenu{\end{enumerate}}
\def\be{\begin{equation}}
\def\ee{\end{equation}}
\def\ba{\begin{eqnarray}}
\def\ea{\end{eqnarray}}
\def\nn{\nonumber}            
\def\R{{\hbox{{\rm I}\kern-.2em\hbox{\rm R}}}}   
\def\H{{\hbox{{\rm I}\kern-.2em\hbox{\rm H}}}}   
\def\N{{\hbox{{\rm I}\kern-.2em\hbox{\rm N}}}}   
\def\C{{\ \hbox{{\rm I}\kern-.6em\hbox{\bf C}}}} 
\def\Z{{\hbox{{\rm Z}\kern-.4em\hbox{\rm Z}}}}   
\def\downarcfill{${\mathsurround=0pt }
\braceld\leaders\vrule\hfill\bracerd $}
\def\arc#1{\mathop{\vbox{\ialign{##\crcr\noalign{\kern3pt}
\downarcfill\crcr\noalign{\kern3pt\nointerlineskip}
$\hfil\displaystyle{#1}\hfil$\crcr}}}\limits}

\def\mettresous#1\sous#2{\mathrel{\mathop{\kern0pt #2}\limits_{#1}}}
\def\sqr#1#2{{\vcenter{\vbox{\hrule height.#2pt
          \hbox{\vrule width.#2pt height#1pt \kern#1pt
           \vrule width.#2pt}
           \hrule height.#2pt}}}}
\def\dalamb{\mathchoice\sqr68\sqr68\sqr{4.2}6\sqr{3}6}
\def\ket#1{|#1\rangle}      
\def\bra#1{\langle #1|}     
\def\kvac {|0\rangle}                   
\def\bvac{\langle 0|}       
\def\braket#1#2{\mathrel{\langle #1|#2\rangle}}   
\def\elematrice#1#2#3{\langle #1|#2|#3 \rangle}   
\def\inoutexpect#1{\elematrice{0,\mbox{out}}{#1}{0,\mbox{in}}}
\def\inout{\langle 0,\mbox{out}|0,\mbox{in}\rangle}

\newlength{\fleche}
\newcounter{FLECHE}
\newcommand{\vecteur}[1]
{
\{
\settowidth{\fleche}{$#1$}
\setcounter{FLECHE}{\fleche}
\setlength{\unitlength}{1sp}
\stackrel
{\begin{picture}(\value{FLECHE},0)
\put(0,0){\vector(1,0){\value{FLECHE}}}
\end{picture}
}
{#1}
\setlength{\unitlength}{1pt} }

\newcounter{DEMIFLECHE}
\newcommand{\lrvect}[1]
{\settowidth{\fleche}{$#1$}
\setcounter{FLECHE}{\fleche}
\newcount\demifleche
\demifleche=\value{FLECHE}
\divide\demifleche by 2
\setcounter{DEMIFLECHE}{\demifleche}
\setlength{\unitlength}{1sp}
\stackrel
{\begin{picture}(\value{FLECHE},0)
\put(\value{DEMIFLECHE},0){\vector(1,0){\value{DEMIFLECHE}}}
\put(\value{DEMIFLECHE},0){\vector(-1,0){\value{DEMIFLECHE}}}
\end{picture} }
{#1} \setlength{\unitlength}{1pt} }

\def\lrpartial{\mathrel{\partial\kern-.75em\raise1.75ex\hbox{$\leftrightarro
w$}}}
\def\lrD{\mathrel{{\cal D}\kern-.75em\raise1.75ex\hbox{$\leftrightarrow$}}}
\def\mettresous#1\sous#2{\mathrel{\mathop{\kern0pt #2}\limits_{#1}}}
\def\zentier{\ \hbox{{\rm Z}\kern-.4em\hbox{\rm Z}}}
\def\reel{\ \hbox{{\rm R}\kern-1em\hbox{{\rm I}}}}
\def\bfGamma{\ \hbox{{$\Gamma$}\kern-.5em\hbox{{\rm I}}}\,}
\def\Iden{\ \hbox{{1}\kern-.25em\hbox{{\rm I}}}\,}
\def\souligne#1{$\underline{\smash{\hbox{#1}}}$}
\def\1op{\ \hbox{{\rm 1}\kern -0.23em\hbox{{\rm I}}}}
\def\ov{\overrightarrow}
\def\spind{\vcenter{\hsize=1em \baselineskip 0pt \parindent 0pt
$\to$ \\ $O$ \\ $\to$}}
\def\spindo{\vcenter{\hsize=1em \baselineskip 0pt \parindent 0pt
$\to$ \\ $\omega$ \\ $\to$}}
\def\spindi{\vcenter{\hsize=1em \baselineskip 0pt \parindent 0pt
$\to$ \\ $I$ \\ $\to$}}
\newcommand{\figdef}[4]
{\begin{figure}
\vglue10pt
\vspace{#1}
\special{illustration #2 scaled 800}
\caption{{\sl #3}}
\vspace{3mm}
\label{#4}
\end{figure}
}
\catcode`\@=11
\@addtoreset{equation}{section}
\def\theequation{\arabic{section}.\arabic{equation}}
\def\appendix{\renewcommand{\thesection}{\Alph{section}}\setcounter{section}{0}
              \renewcommand{\theequation}
{\mbox{\Alph{section}.\arabic{equation}}}\setcounter{equation}{0}}
\renewcommand{\thesection}{\Roman{section}}
\renewcommand{\thesubsection}{\thesection-\Alph{subsection}}

\vskip 1. cm
\begin{center} {\LARGE\bf   Decreasing { \em circumference }  for increasing
{ \em radius } in axially symmetric gravitating systems }\\
\vskip 1. cm M. Lubo\footnote{E-mail muso@umhsp02. umh. ac.
be}$\mbox{}^{\mbox{\footnotesize }}$,
\\ {\em M\'ecanique et Gravitation, Universit\'e de Mons-Hainaut},
\\ {\em 6, avenue du Champ de Mars, B-7000 Mons, Belgium}
 \\
\vskip 0.5 cm
\vskip 1 cm
\end{center}
\begin{abstract}
Apart from the flat space with an angular deficit, Einstein
general relativity possesses another cylindrically symmetric
solution. Because this configuration displays circles whose
``circumferences'' tend to zero when their ``radius'' go to
infinity, it has not received much attention in the past. We
propose a geometric interpretation of this feature and find that
it implies field boundary conditions different from the ones found
in the literature if one considers a source consisting of the
scalar and the  vector fields of a $ U(1) $ system . To obtain a
non increasing energy density the gauge symmetry must be unbroken
. For the Higgs potential this is achieved only with a vanishing
vacuum expectation value but then the  solution has a null scalar
field. A non trivial scalar behaviour is exhibited for a potential
of sixth order. The trajectories of test particles in this
geometry are studied, its causal structure discussed. We find that
this bosonic background can support a normalizable fermionic
condensate but not such a current.
\end{abstract}
\newpage
\section{Introduction}
\par Among the most important characteristics of cosmic strings is the
existence of a symmetry axis and the concentration of energy
around this axis \cite{LABEL1}. One can ask if cosmic strings are
the only configurations displaying this feature.  Taking into
account gravity, the existence of a symmetry axis implies
cylindrical symmetry for the metric as well. All static
cylindrically symmetric solutions of Einstein equations in vacuum
are of the Kasner type. The vanishing of the energy-momentum
tensor in the ``asymptotic'' region of any physically reasonable
solution implies that the geometry should approach a Kasner line
there. As the energy momentum tensor corresponding to these axial
configurations implies the invariance of the metric under boosts,
one is left with only two Kasner geometries : a flat space
presenting a conical singularity and a Melvin like space
\cite{LABEL2}. The first case leads to the well known cosmic
strings whose possible role in cosmology ranges from structure
formation to baryogenesis. The second solution has received less
attention because it seems to present a singular behaviour in the
metrical sector : circles of increasing ``radius'' display
decreasing circumferences. Recent investigations suggest that for
an Abelian-Higgs system with a symmetry breaking scale much
smaller than the Planck scale,  asymptotically Melvin-like
solution coexist with the habitual string configuration as long as
the angular deficit of the latter does not exceed $2 \pi$
\cite{LABEL3}.

\par The inertial and Tolman masses of the two  solutions have been
contrasted \cite{LABEL4}. However, the unusual behaviour of the
metric has not been addressed. The key point of this article is
the interpretation of this unusual geometry. We shall argue that
the coordinate which has been considered as the radius in the
second special Kasner-line element (i.e. the Melvin-like geometry)
is a rather complicated function of the ``true'' radius. To
support our proposal, we shall invoke the coordinatization of the
sphere by polar stereographic variables in the first section. This
example, although elementary, strongly suggests that letting $r$
go to infinity in the second special Kasner line element, one
enters an asymptotic region  {\sl {which shrinks  to the symmetry
axis}}. In the third section, an Abelian Lagrangian is coupled to
the Einstein-Hilbert one. Looking for an axially symmetric
configuration different of the string, the preceding section
imposes the vanishing of the vector and the scalar field as the
coordinate $r$, formerly interpreted as a radius , goes to
infinity. This is at odds with previous treatments and results in
a divergent inert mass per unit length if the v.e.v of the Higgs
field does not vanish for example. In the fourth section we find
that the trajectories of massive particles are bounded from above
in this space time. We also study massless particle trajectories
and discuss the causal structure of the new solution. Due to the
cylindrical symmetry, we propose a Penrose- like diagram, but with
each point representing a two geometry conformal to a  cylinder.
In the fifth section, we study the wave function which describes a
condensate or a current of a fermion charged under the Abelian
field considered.

\section{The second special Kasner line element.}
The metric
\be
 \label{1} ds^2 = c \, r^{4/3} (dt^2-dz^2)-dr^2 -  d
\,r^{-2/3}d\theta^2
\ee
where $c$ and $d$ are positive constants
is what we call the second Kasner line element. It has been shown
to satisfy Einstein equations in the vacuum \cite{LABEL2} ; it
displays a cylindrical symmetry if $\theta=0$ is identified with
$\theta=2\pi$.
\par If one considers the circle parameterized by $t=t_0, z=z_0,
r = r_0$ and $\theta \in [0, 2\pi]$, its circumference $2\pi
r_0^{-2/3}$ decreases as the ``radius'' $ r_0$ increases. But, the
circle given by $ r_0 = \infty$ being space-like everywhere and
having a null length can not be anything else than a point.
\par This is already realized on the sphere. Consider a point $Q(x,y,\sqrt{ \sigma^2 - x^2 -y^2})$
located on the sphere of radius $\sigma$ centered at
$(0,0,\sigma)$. The straight line joining $Q$ to the north pole
intersects the plane $z=0$ at the point $P(a,b,0)$ specified by
\be
 \label{2}
 x=a(1-\lambda)\qquad y = b(1-\lambda) \ee where \be
\label{3} \lambda = {2(a^2+b^2) + \sigma^2+\sigma
\sqrt{4\sigma^2+3(a^2+b^2)}\over {2(a^2+b^2+\sigma^2)}}.
\ee
Introducing the polar stereographic coordinates ($ r, \theta$) by
$a = r \cos \theta$, $b = r \sin \theta$ one obtains
 \be \label{4}
ds^2 = {16\sigma^4\over {(4\sigma^2+ r^2)^2}} dr^2 + {16 \sigma^4
r^2\over {(4\sigma^2+ r^2)^2}} d\theta^2. \ee The coefficient
$g_{\theta\theta}$ becomes a decreasing function for large values
of the coordinate $r$. If one does not know where this metric
comes from and interpret it  as a radius, a circle of infinite
radius turns out to be of null length. This is not surprising
since $r = \infty$ corresponds to the point at infinity on the
plane which is mapped into the north pole by the  stereographic
projection. Here , $ r = \infty $ is just the north pole. When the
coordinate $ r$ vanishes, one has another circle displaying a
vanishing circumference : the south pole. The two are on the
symmetry axis. Introducing the variable \be \label{5} r_{\ast} =
2\sigma \arctan (r/2\sigma) \ee
 the metric reads
\be \label{6} ds^2 = dr^2_{\ast} + \sigma^2 \sin^2
\left({r_{\ast}\over {\sigma}} \right) d\theta^2. \ee The relation
between $ r_{\ast}$ and $r$ is bijective provided that $r_{\ast}
\in [ 0, \pi \sigma ]$. The points located on the symmetry axis
once again are those for which the coefficient $g_{\theta\theta}$
vanish.
\par The section $t = c^{st}$ , $z= c^{st} $
 of the second Kasner line element
\be \label{7} ds^2  = dr^2 + b r^{-2/3} d\theta^2 \ee is a two
dimensional surface ; the preceding discussion strongly suggests
that $r = \infty$ is a ''point'' on the symmetry axis. The
difference with the sphere lies in the fact that the second
special Kasner geometry is unbounded since the meridian $\theta =
c^{st}$ is of infinite length; $ r = \infty $ is a point rejected
at infinity on the symmetry axis.
\par This interpretation is also supported by an embedding diagram. The
previous two geometry can be rewritten as
 \be
 \label{8} ds^2 = 9
b^{-3} \rho^{-8} d\rho^2 + \rho^2 d\theta^2.
\ee by the trivial
change of coordinates $ r = (b^{-1}\rho^2)^{-3/2}$. This line
element  can be realized in the Euclidean space as the surface of
revolution
 \be
 \label{9} z(\rho) = \int^{\rho}_0 d\xi (\rho^8_{cr}
 \xi^{-8} -1)^{1/2}. \ee which is real only for the values of $
\rho $ satisfying the inequality
\be
 \label{10} \rho \leq
\rho_{cr} = (9 b^{-3})^{1/8}.
 \ee This can be translated for the
geometry given in Eq. (\ref{8}) by  $r \geq (b
\rho^{-2}_{cr})^{3/2}$. The coordinate $r$ measures the length
along a generating line while $ \rho $ is linked to the
circumference by the usual formula $ 2 \Pi \rho $ . The two
geometry looks like a  bottle with an infinite neck which becomes
thinner and thinner ( as $ r -> \infty, \rho -> 0 $)  and a basis
which becomes larger and larger(as $  r -> 0 , \rho  -> \infty $)
. The region $r \simeq 0$ (where the components of the Riemann
tensor associated to geodesic deviations become infinite) is
excluded from the embedding and will be excluded when
 raccording the second Kasner line element to non vacuum solutions of Einstein equations.
 Our embedding is similar to the one   developed  in the case of the Melvin solution
  \cite{LABEL5}. The new feature here is that this fact will be used with the  example of the sphere to argue that new boundary conditions arise.
\par  What is the causal structure of this second
Kasner special geometry? As the relevant symmetry here is axial, we will
factorize the geometry of a cylinder. Introducing a radial coordinate
$\rho$ (different from the one appearing in Eq. (\ref{8})) by

\be \label{11} r = b^{1/2} a^{-1/2} \rho^{-1} \ee one obtains
 \be
\label{12} ds^2 = a^{-1/3} b^{2/3}  \rho^{-4/3}[-dt^2+
a^{-4/3}b^{1/3} \rho^{-8/3}d\rho^2]+ a^{-1/3} b^{2/3} \rho^{-4/3}
[dz^2+\rho^2d\theta^2]. \ee In a spherically symmetric geometry
like the Schwarzchild solution, the causal structure is found by
making the coordinates $\theta$ and $\phi$, which parameterizes a
sphere, constants. The remaining geometry is written in terms of
bounded ingoing and outgoing null coordinates \cite{LABEL6}. In
the Penrose diagram, each point represents a sphere. Following
this procedure, we shall take $z$ and $\theta$ ( which
parameterizes the cylinder) to be constants in Eq.(\ref{12}) :
this gives the light-cone coordinates \be \label{13}
\left.\begin{array}{ll}
\bar u\\
\bar v
\end{array}\right\rbrace
= \pm t-3 a^{-2/3} b^{1/6} \rho^{-1/3} \ee in terms of which the
metric reads
 \be
 \label{14} ds^2 = -{1\over 6} a^{2/3} b^{5/6}
{(\tan\bar u + \tan \bar v) \over {\cos^2\bar u \cos^2\bar v}}
d\bar u d\bar v.
 \ee
 In a conformal diagram  , the region $A$
specified by ( $r = 0$ , $t$ finite ) would be  a singular region(
The Gauss scalar curvature diverges there)  while the ``region''
$B$ given by ( $ r = \infty$, $t$ finite ) would correspond simply
to a point. It should be remembered that each point on such a
diagram is a surface conformal to a cylinder. The difference with
the spherical solutions lies in the fact that here we probe the
space-time structure using horizontal light rays which cross the
symmetry axis  at right angles while in the former one uses radial
ones. The intermediary variable $ \rho $ was only illustrative
since Eq.$( \ref{13} )$ and Eq.$( \ref{11}) $ give the link
between the null coordinates and the set of variables $ t,r$ . We
shall omit this step in the coming section.
\par We  shall be interested in the causal structure of a solution of the gravitating
Abelian system which has the same asymptotic behavior than the
Kasner geometry.

\section{The gravitating $U(1)$ system.}

\par Let us consider a self gravitating Abelian-Higgs system minimally
coupled to gravity. The classical field equations are derived from
the action
\be
\label{15}
S=\int d^4x \sqrt{-g}\left[{1\over 2} D_{\mu} \Phi D^{\mu} \Phi^{\ast}-
{\lambda\over 4} (\Phi \Phi^{\ast} - v^2)^2 - {1\over 4} F_{\mu\nu} F^{\mu\nu}
+{R\over {16\pi G}}\right].
\ee
The $U(1)$ charge $e$ is embodied in the covariant derivative
\mbox{ $D_{\mu}\Phi = \partial_{\mu}\Phi - i eA_{\mu}\Phi$ }. For a static
cylindrically symmetric configuration, the ens\"atze can be given the
form
\be
\label{16}
ds^2 = N^2(r) dt^2-dr^2 - L^2(r) d\theta^2-K^2(r)
dz^2
\ee
\be
\label{17}
\Phi =  f(r) e^{i\theta} \quad;\quad A_{\mu}dx^{\mu} = e^{-1}(1-P(r))
d\theta.
\ee
\par The field's coordinate dependence  leads to the equality
$T^0_0 = T^z_z$ in the entire space-time and so implies $N(r) =
K(r)$. Among all the Kasner geometries, this condition selects the
cosmic string solution (a flat geometry with an angular deficit)
and the second Kasner line element given in Eq.(\ref{1}). The
cosmic string solution has been extensively studied in the
literature. In that configuration, the smoothness of the geometry
on the symmetry axis is guaranteed by the initial conditions
\cite{LABEL2} \be \label{18} L(0) = 0\ ,\ L'(0) = 1\ ,\ K(0) = 1\
,\ K'(0) = 0 \ee while the matter fields are non singular on the
core provided that \be \label{19} f(0) = 0\ ,\ P(0) = 1 . \ee
\par The finiteness of energy implies
\be \label{20} f(\infty) = v \quad , \quad P(\infty) = 0. \ee What
happens if one considers the second special Kasner line element of
Eq.(\ref{1}) as giving the asymptotic behavior  of the metric i.e.
when the coordinate $r$ goes to infinity? In the previous section,
we argued that $ r = \infty$ is the point at infinity on the
symmetry axis. For a regular configuration, the Higgs and the
vector fields must vanish there \be \label{21} f(\infty) = 0\quad
and \quad P(\infty) = 1. \ee
\par Extracting the expression of the inertial mass from Eq.(\ref{15}) one
has \be \label{22} E = \int dr d\theta dz  LK^2 \left[{1\over 2}
g_{rr} |D_r \Phi|^2 + {1\over {2 }} g_{ \theta \theta} |D_\theta
\Phi|^2
 + {1\over 4} F_{r
\theta} F^{r \theta} + {\lambda\over 4} (\Phi^{\ast} \Phi- v^2)^2
\right]. \ee In the asymptotic region (i.e. $ r \rightarrow
\infty$) one has $L K^ 2  \sim r$ so that the volume element is
not bounded. The first three terms decrease in the asymptotic
region provided that $f(r)$ and $P(r)$ approach constants there;
this is already satisfied by Eq.(\ref{21}). The contribution of
the Higgs potential in this part of the space is   reduced to the
integral of  $ r \lambda v^4 $. This has a chance to converge
 only when $v=0$ i.e. when the system does not face a spontaneous symmetry breaking,
(at least at tree level). The vanishing of the v.e.v makes it
impossible to use the habitual
 parameterization $ \Phi = v f(x) e^{ i \theta} $ in terms of  the dimensionless length $ x =
 \sqrt{ \lambda v^2} r $. Nevertheless, the dimensionfull Newton constant $G$ makes possible
  the parameterization $ \Phi = (1/\sqrt{ G \lambda} ) f(x) e^{ i \theta} $ in terms of the
 dimensionless quantity
  $ x = \sqrt{G} r$ measuring the length. In the same spirit, we  can introduce other
  dimensionless
   functions
  $  P(x) , K(x), L(x)$ and  write
   the field equations obtained by extremizing
the action given in Eq.(\ref{15}). In this case, we obtained a
vanishing scalar field for any value of the parameters. Imposing
the boundary conditions specified by  Eq.$(\ref{19}) $ and
Eq.$(\ref{21})$, one obtains a function $ \bar{f(x)} $ whose
values are many orders of magnitude smaller than the error fixed
for numerical computations. This has to be discarded. Physically
this can be understood as follows. Forcing the scalar field to go
from zero to zero as $r$ goes from  zero to infinity, one obtains
that it vanishes identically since there is no source. Such a
source would be for example a local maximum of the potential but
as the vacuum expectation value vanishes, such a maximum does not
exist. Taking an identically vanishing scalar field, we found a
solution. In order to obtain a smooth embedding diagram, we took
the coordinate $r$ to range from $ - \infty $ to $ \infty $. This
is realized simply by taking $r$ to be the oriented length on a
meridian of the two surface $ t = c^{st} , z = c^{st} $ : above
the equator $r$ is taken positive and below it, it becomes
negative. To ensure a symmetric embedding , we choosed the
conditions \be \label{27}
 f'(0) = P'(0) = L'(0) = K'(0) = 0
\ee while at spatial infinity( $ r = \pm \infty $), the boundary
conditions displayed in  Eq. $ ( \ref{21})  $ were imposed.  To
give a unified description with the case where a non vanishing
scalar field is present, the equations of motion are given below (
Eq.$( \ref{23})$- Eq.$( \ref{26})$ ) in the case of an appropriate
potential. .
   \par Is it possible to construct a solution with a non vanishing scalar field? To do this we
need a potential which vanishes with the scalar field so that the
minimum is attained at spatial infinity. We also need a local
maximum which will correspond to a source. These conditions are
for example satisfied by the gauge invariant potential \be
\label{28}
  V( \Phi) = \frac{\lambda}{v^2}  \Phi \Phi^{*} (  ( \Phi \Phi^{*}) -  v^2
  )^2
  \ee

 $v$ being an inverse length scale. A local maximum is attained
at $ | \phi| = v/\sqrt{3}  $ while there are  two  minima , at  $
| \phi | = 0 , v $. The U(1) symmetry is broken spontaneously in
the second vacuum and preserved in the first one. We disregard the
 renormalizability since we are interested only in classical solutions. Looking for a
 configuration whose
 geometry has an embedding similar to the one constructed with a vanishing scalar field, the
 radius $ r= 0$ is particular because it corresponds to the greatest
circumference as can be seen from $  Eq.(\ref{27}) $ . Here one
uses the usual parameterization $ r = \sqrt{\lambda v^2} x $ , $
\Phi = v f(x) \exp{ i \theta} $ , $ \cdots $ of \cite{LABEL3} and
the dimensionless quantities $ \alpha = e^2/\lambda , \gamma = 8
\pi G v^2 $ . The Euler Lagrange equations read
 \be
 \label{23}
f''(x) + \frac{ f'(x) L'(x)}{L(x)} + 2 \frac{ f'(x) K'(x)}{K(x)} -
  \frac{ f(x) P(x)^2}{L(x)^2} - 6 f(x)^5 + 8 f(x)^3 - 2 f(x) = 0
\ee

\be \label{24} P''(x) - \frac{ L'(x) P'(x)}{L(x)} + \frac{
      K'(x) P'(x)}{K(x)} - \alpha f(x)^2  P(x) = 0
\ee

\begin{eqnarray}
\label{25} K''(x) & - & \gamma \frac{K(x) P'(x)^2}{2 \alpha
L(x)^2} + \frac{
      K'(x)  L'(x)}{L(x)} + \frac{ K'(x)^2}{K(x)}  \nonumber\\
      & + & \gamma   f(x)^6   K(x) -
  2  \gamma   f(x)^4  K(x) + \gamma f(x)^2   K(x) = 0
\end{eqnarray}

\begin{eqnarray}
\label{26} L''(x)  & + &  \gamma \frac{P'(x)^2}{2  \alpha L(x)} +
  2  \frac{ K'(x) L'(x)}{K(x)} +
  \gamma   \frac{f(x)^2  P(x)^2}{L(x)}  +    \gamma   f(x)^6   L(x) \nonumber\\
   & - &
  2  \gamma   f(x)^4  L(x) + \gamma f(x)^2   L(x) = 0
\end{eqnarray}

The  fields dependence of this  solution are plotted in Fig.1.1 -
Fig.1.4 . for the parameters $ \alpha =  1 , \gamma  = 0.01$. The
scalar is the most rapidly varying field,
  followed by the vector field. The components of the metric
  change significantly on a scale approximately equal to ten times
  the one needed for the matter fields. The energy density per unit length along $ z $ , noted $
\epsilon (x)$ and defined by $ E = 2 \pi v^2  \int dx  \epsilon
(x) $ is plotted in Fig1.5 . It has significant contributions at $
x = \pm 1.39 $ where the scalar
  field attains the value corresponding to the maximum of the
  potential.
  An asymptotic analysis shows the total energy converges. For our
  example, $ E = 2 \pi v^2  0.79 $ .

\section{ The classical trajectories.}
Let us first consider the case of  the massive particles. The
metric specified in Eq.(\ref{16}) has three cyclic coordinates ;
its geodesics are by way of consequence characterized by three
constants of motion $E, l, a$ such that $ N^2(r)\dot t = E, L^2(r)
\dot \phi = l , K^2(r)\dot z=a $. The derivatives are performed
with respect to the proper time of the observer. Introducing the
proper time per unit mass $\tau$, the energy-momentum relations
reads \be \label{31} \left( {dr\over{d\tau}} \right)^2 =
{E^2-a^2\over {\mu^2}} {1\over {K^2(r)}} -{l^2\over {\mu^2}}
{1\over {L^2(r)}} - 1\equiv A(r) \ee $\mu$ being the mass of the
particle. A physical motion possesses a real velocity. A necessary
condition for this to be possible is $E^2>a^2$. The asymptotic
behaviour of the metric displayed in Eq.(\ref{1}) shows that the
trajectory of a massive particle is always bounded from above :
such a particle will never reach the point at infinity on the
symmetry axis( $ r = \infty $ ).

The physically allowed region exhibits positive values of $A(r)$.
One of the signatures of this geometry is the existence of a
maximum of the function $L(r)$ which means that the set of slices
$t$=constant, $z$= constant possess a maximal circumference. The
angular velocity of an observer attains its lowest value there.

\par  The causal Structure is found by analyzing particular light geodesics.
The bounded null coordinates in the new background are given by
$\bar u = c^{st}$ or $\bar v = c^{st}$ with \be \label{33}
\left.\begin{array}{l}
\bar u \\
\bar v
\end{array}\right\rbrace = \arctan [t \mp \sigma(r)]
\ee
  with
\be
\label{34}
   \sigma(r) = \int_0^r d\xi \frac{1}{ K(\xi)}
\ee
 The $2$ metric
obtained by fixing $\theta$ and $z$ can now be written as
\be
\label{35} ds^2 = \frac{d\bar u d\bar v }{\cos^2 \bar u \cos^2
\bar v}   K \left(  \sigma^{-1}
 \left( \frac{ \tan \bar u -  \tan \bar v}{2}  \right) \right)
\ee
\par For the embedding of this two geometry in the  Euclidean space, let us consider
the change of
the coordinate variable $\rho = L(r)$ which is valid only above or
below the radius $r = 0 $ which maximizes $L(r)$ . The  two
dimensional geometry
 \be
\label{36} ds^2 = L^2 (r) d \theta^2 +  dr^2
\ee
can be realized
in an Euclidean 3 space as the surface of revolution \be
\label{37} z(\rho) = \int^{\rho}_{\rho_0} d\xi \left[
\frac{1}{L'(L^{-1}(\xi))^2} - 1 \right]^{1/2}. \ee This is
possible only in the regions in which the term under the square
root is positive.

The embedding of the slice $ t = c^{te}, z = c^{te}$ is generated
by the rotation of the curve displayed in Fig.3 around the $ y$
axis  ; it looks like a bottle with two necks. The associated
causal structure is given in Fig.3.

\section{ The coupling to a fermionic field.}
   In the background of many configurations defined by bosonic fields, it proves
useful to look for fermionic solutions. For example, the sphaleron
possesses a fermionic charge \cite{LABEL7} and a cosmic string can
support a fermionic current \cite{LABEL8} .
\par Let us consider a fermion charged under the gauge field. As the gauge symmetry is manifest,
it has to be massless. We shall write its wave equation in the bosonic background studied above.
Our choice for the $\gamma$
matrices is the following
\be
\label{38}
\gamma^t = K(r)^{-1}
\left(\begin{array}{cccc}
1 &0&0&0\\
0 &-1&0&0\\
0 &0 &-1&0\\
0 &0&0&1
\end{array}\right)
\qquad
\gamma^r =
\left(\begin{array}{cccc}
0 &e^{-i\theta}&0&0\\
-e^{i\theta} &0&0&0\\
0 &0 &0&-ie^{-i\theta}\\
0 &0&e^{i\theta}&0
\end{array}\right)
\ee

\be
\label{39}
\gamma^0 = L(r)^{-1}
\left(\begin{array}{cccc}
0 &-ie^{-i\theta}&0&0\\
-ie^{i\theta} &0&0&0\\
0 &0 &0&ie^{-i\theta}\\
0 &0&ie^{i\theta}&0
\end{array}\right)
\qquad
\gamma^z = K(r)^{-1}
\left(\begin{array}{cccc}
0 &0&1&0\\
0 &0&0&1\\
-1&0 &0&0\\
0 &-1&0&0
.
\end{array}\right)
\ee
The ens\"atze we shall use for the Dirac field is
\begin{eqnarray}
\label{40} \psi^t & = &  (  \psi_1(r)\exp{(i(k z-\omega t + n_1
\theta))}, \psi_2(r)
\exp{(i(k z-\omega t + n_2 \theta))}  \nonumber\\
& & \psi_3(r) \exp{( i(k z-\omega t + n_3\theta)) }, \psi_4(r)
\exp{( i(k z-\omega t + n_4 \theta)}) ) .
\end{eqnarray}
Factorizing the dependence in $\theta$ one finds the relations \be
\label{41} n_2=n_1+1, n_3=n_2-1, n_4=1+n_1 \ee while the
reparameterization \be \label{42} \psi_1(r) = i \chi_1(r)\ ,\
\psi_2(r) = \chi_2(r)\ , \ \psi_3(r) = i \chi_3(r) \ ,\ \psi_4(r)
= \chi_4(r) \ee renders the equations purely real. We shall
specialize to two possibilities which have been analyzed for the
cosmic string \cite{LABEL8}.

\subsection{ The condensate}

Assuming $k=\omega=0$, one is left with the two decoupled
equations
\be
\label{43}
\chi'_1(r) -{(-1+n_1+P(r))\over{L(r)}} \chi_1(r) = 0
\ee
\be
\label{44}
\chi'_2(r) + {(n_1+P(r))\over{L(r)}} \chi_2(r) = 0
\ee
supplemented by the equalities
\be
\label{45}
\chi_3(r) = \chi_1(r)\ ,\ \chi_4(r) = \chi_2(r).
\ee
\par The solution of the second differential equation can be obtained
using quadratures :
\be
\label{46}
\chi_{2 P}(r) = \exp \left(-\int^{r}_{r_0} {(n_1+P(\xi))\over {L(\xi)}}
d\xi\right)
\ee
$ r_0$ being an arbitrary constant .\\
As $P(\xi)\sim 0$ and $L(\xi)\sim \xi^{-1/3}$ in the asymptotic region,
one obtains
\be
\label{47}
\chi_{2 P}(r) \sim c^{st} \exp \left(-{3\over 4} (n_1 + 1) r^{4/3} \right)
\ee
which is normalizable only if $n_1> -1$ since the volume element yields a factor
$ r^{1/3} $.

\par Similar considerations give
\be \label{48} \chi_{1 P}(r) \sim \exp \left( {3\over 4} n_1
r^{4/3}  \right)
 \ee
which is normalizable only when $ n_1 < 0 $.

 To summarize, the wave functions
\be \label{49} \psi^t = i \chi_{1 P}(r) e^{i n_1\theta} (1,0,1,0)
\qquad    \psi^t = i \chi_{2 P}(r) e^{i n_1\theta} (0,1,0,1) \ee
with  $ n_1 < 0 $ in the first case and $ n_1 < -1 $ in the second
case are normalized and describe  condensates.

\subsection{ The current.}

The dispersion relation $k=\omega$ preserves the equalities $\chi_4 =
\chi_2, \chi_3 = \chi_1$ but changes the field equations into
\be
\label{50}
\chi'_1(r) -{(-1+n_1+P(r))\over {L(r)}} \chi_1(r)
- {2\omega\over{K(r)}} \chi_2(r)
\ee
\be
\label{51}
\chi'_2(r) + {(n_1+P(r))\over {L(r)}} \chi_2(r)=0.
\ee
\par The last equation is exactly  one of those  found for the
condensate so that we can take $ \chi_2(r) = \chi_{2P }(r) $ with
$ n_1 > - 1 $. The remaining equation is solved by

\be
\label{52}
  \chi_1(r) =  \chi_{1 P}(r) \int_{ro}^r d\xi \xi^{-2/3} e^{ - {3
  \over 4} (2 n_1 + 1) \xi }.
\ee
  The integral present in this formula converges only for the integers $ n_1$
  satisfying $
  n_1 > - {1\over 2} $. But for these values, the function $   \chi_{1
  P}(r)$ goes to infinity so that the wave function is not
  normalizable.

\be
\label{55}
\chi'_1(r) - {(-1+n_1+P(r))\over L(r) } \chi_1(r) = 0
\ee

\be \label{56} \chi'_2(r) + {(n_1+P(r))\over {L(r)}} \chi_2(r) +
{2\omega\over {K(r)}} \chi_1(r) = 0 . \ee The solution to the
first equation has been seen to be well defined when $ n_1 < 0 $.
Calling $\chi_{2P}(r)$ the solution of  Eq.(\ref{43}), one finds
that \be \label{57} \chi_2(r) = -\chi_{2P}(r) \int^{r}_{r_0}
{2\omega \over {K(\xi)}} {\chi_1(\xi)\over{\chi_{2P}(\xi)}} d\xi
\ee solves Eq.(\ref{50}). However, the behaviour of $\chi_1(\xi)$
and $\chi_{2P}(\xi)$ in the asymptotic region makes this
$\chi_2(r)$ diverge.

\section{ Conclusion }
\par In the asymptotic region, the geometry we studied is close to
the one exhibited by the Melvin solution and the ones constructed
recently \cite{LABEL3, LABEL4} , but the magnetic field looks
quite different.
\par The mechanism for the formation of this
configuration has not been addressed. In particular, one can ask
if like monopoles or domain
 walls, it does not come to dominate the energy density of the universe. This may be used to
 constrain the parameters of the model.
\par A section $ t = c^{te}, z = c^{te} $  of the geometry of a
cosmic string is a plane with an angular deficit. Putting such
planes in parallel, with the edges one above the other, one
obtains the representation of the $3$ geometry corresponding to a
fixed time. We could not do the same for the solutions we
constructed. This is an important and open problem.
 \par Hawking et
al. have shown that a generalization of the Melvin solution( the C
metric ) plays a role in the creation of black hole pairs
\cite{LABEL9}. It may prove interesting to study to which extent
their picture is affected by our interpretation.
\par A point of
view different  from the one adopted here would
  consist in taking  the variable $ \theta $ to be a non compact
  coordinate. The configuration displayed in $ Eq.( \ref{1})$ , being a solution depending on a
   sole distance $r$
   may then interpreted as a sort of domain wall. Such a
   possibility  would require a specific treatment and has not been considered here.

\underline{Acknowledgement}
   I thank G.Senjanovic, R.Jeannerot of I.C.T.P ( where a part of this work was performed last
   year), Ph.Spindel and Cl.Gabriel for stimulating discussions. I also thank the F.N.R.S for
   financing my travel to Trieste.

\section{Figures captions}

\begin{itemize}

  \item   Fig1.1 - Fig1.5 displays the fields dependence and the energy density
   of the  solution
  for the parameters $ \alpha = e^2/\lambda = 1 , \gamma = 8 \Pi G
  v^2 = 0.01$.
  \item   Fig.2 gives the generating line of the embedding of the slice $ t = c^{te} ,
  z = c^{te}$ of the
  solution as a surface of revolution  in the Euclidean space .
  The part which hugs the rotation axis has not been rendered with
  precision.
  The rotation is carried around the $y$ axis.
  \item   Fig.3 gives the causal structure of the
  solution. The line $ r = 0 $ is the diagonal ,  the thick line represents a line  $  r = r_o $
     , a the dotted line corresponds  $ t = t_o $ . The time-like
     infinities are $ I^{+} = ( \bar{u} = \bar{v} = \pi/2 ) $ and $ I^{-} = ( \bar{u} = \bar{v}
      = - \pi/2 ) $. As $ r$ can change sign, there are two space
      infinity : $ I^{>}_{o}
= ( - \pi/2 , \pi/2 ) $ and  $ I^{<}_{o} = (  \pi/2 , - \pi/2) $.
\end{itemize}

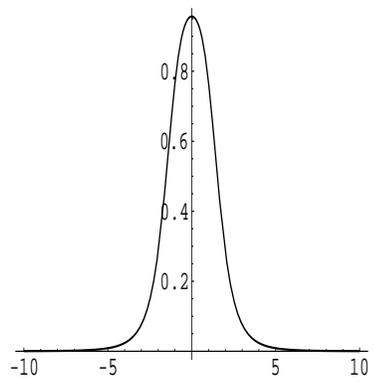
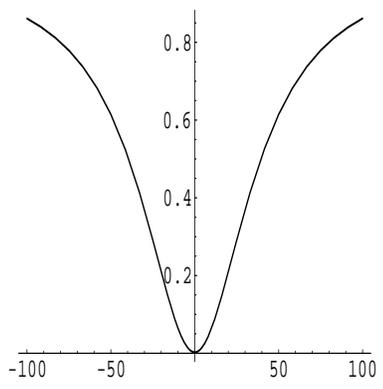
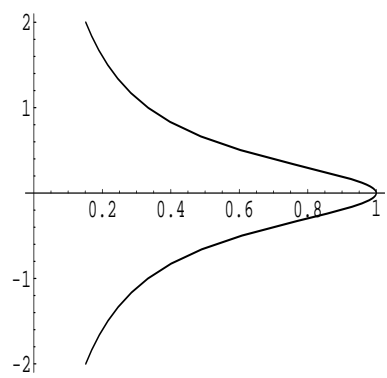
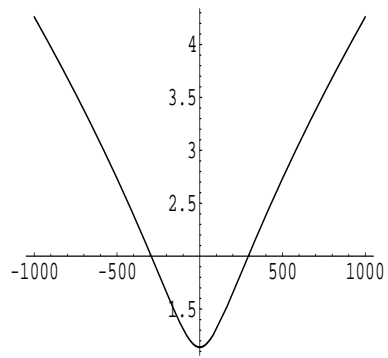
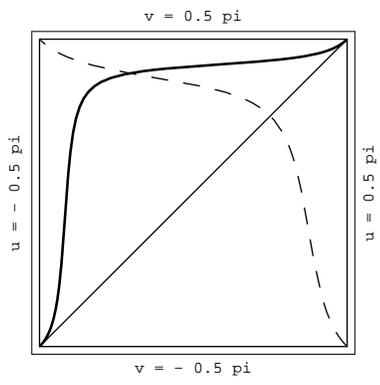
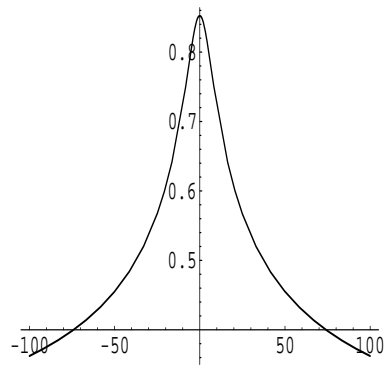
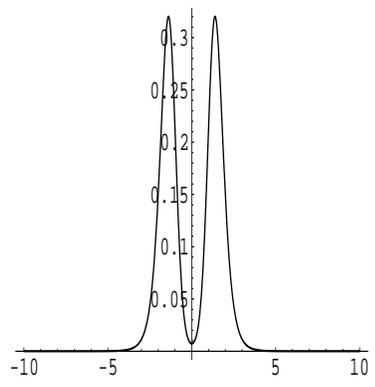
\end{document}